\documentstyle[amstex,amssymb,preprint,aps]{revtex}
\begin{document}
\title{Scalar Gluonium and Instantons\thanks{%
HD-TVP-00-1} }
\author{Hilmar Forkel}
\address{Institut f{\"u}r Theoretische Physik, Universit{\"a}t Heidelberg, \\
D-69120 Heidelberg, Germany}
\date{\today}
\maketitle

\begin{abstract}
The impact of QCD instantons on scalar glueball properties is studied in the
framework of an instanton-improved operator product expansion (IOPE) for the 
$0^{++}$ glueball correlation function. Direct instanton contributions are
found to strongly dominate over those from perturbative fluctuations and
soft vacuum fields. All IOPE sum rules, including the one involving a
subtraction constant, show a high degree of stability and are, in contrast
to previous glueball sum rules, consistent with the low-energy theorem for
the zero-momentum correlator. The predicted glueball mass $m_{G}=1.53\pm 0.2$
GeV is less sensitive to the instanton contributions then the glueball
coupling (residue) $f_{G}=1.01\pm 0.25$ GeV, which increases by about half
an order of magnitude. Both glueball properties are shown to obey scaling
relations as a function of the average instanton size and density.
\end{abstract}


\section{Introduction}

Glueballs, as the most immediate manifestation of gluonic self-interactions
in the hadron spectrum, represent a key challenge for our understanding of
nonperturbative Yang-Mills dynamics. Not surprisingly, therefore,
theoretical interest in gluonium dates back to the early days of QCD \cite
{gel72} and has spurred intense research activity ever since. Estimates of
glueball properties were obtained in a variety of approaches, ranging from
model analyses \cite{mod} and steadily improving lattice simulations \cite
{lat} to QCD sum rule calculations \cite{nov280,nov81,nar84,bag90,nar98}.

The QCD sum-rule technique, in particular, combines the advantages of an
analytical approach with a firm and largely model-independent basis in QCD
and should therefore be well suited for obtaining qualitative and
quantitative insight into the glueball spectrum. During its early
development, however, it became clear that this approach encounters
conceptual and practical problems in the scalar glueball channel \cite
{nov280,nov81}. These problems, which have so far prevented fully consistent
and conclusive sum-rule predictions, can be traced to the exceptionally
strong coupling of the vacuum to the spin-0 glueball interpolators. The
intense vacuum response generates nonperturbative violations of asymptotic
freedom starting from unusually small distances $x\sim 0.02$ fm \cite{nov81}%
, and the power corrections of the conventional operator product expansion
(OPE) are much too weak to account for this physics \cite{nov280}. As a
consequence, stability and mutual consistency of different sum rules (see
below) is partially lacking, and serious difficulties are encountered in
reconciling the sum rules with the low-energy theorem \cite{nov280} which
governs the long-distance behavior of the scalar glueball correlator.

In search of the origin for the apparently missing nonperturbative
short-distance physics it is natural to expect that it may at least partly
be provided by small (or ``direct'') instantons \cite
{nov280,nov81,shu82,sch95}. Indeed, as coherent vacuum gluon fields
instantons couple particularly strongly to the gluonic interpolators of the $%
0^{++}$ glueball channel. Nevertheless, they are neglected in the
perturbatively calculated Wilson coefficients of the conventional OPE. An
additional and\ perhaps more intuitive reason for expecting instanton
physics to play a major role in the structure of scalar glueballs derives
from their exceptionally small size $r_{G}\simeq 0.2$ fm, found in lattice 
\cite{def92} and instanton vacuum model \cite{sch95} calculations. Since $%
r_{G}$ is much smaller than the confinement scale and the size of heavier
glueballs, it is rather unlikely that scalar glueballs are bound
predominantly by (iterated) perturbative or by confinement forces. The
attractive interactions mediated by instantons provide a suggestive
alternative.

Motivated by the above considerations, our main objectives in this paper are
to evaluate the direct instanton sector of the scalar glueball correlator by
means of an instanton-improved OPE (IOPE) and to analyze the ensuing
glueball sum rules. While analytical instanton calculations (at low
energies) have long been hampered by insufficient knowledge of the instanton
size distribution and notorious infrared problems, this impasse can nowadays
be avoided by relying on the results of instanton vacuum model \cite{sch98}
and lattice \cite{instlat} simulations for the required bulk features of the
instanton distribution, i.e., for the average instanton size $\bar{\rho}%
\simeq (1/3)$ fm and density $\bar{n}\simeq (1/2)$ fm$^{-4}$. Despite
remaining numerical discrepancies pertaining to the distribution of
large-size instantons, these scales provide\ a solid foundation for our
calculations below to which only small instantons of sizes $\rho \lesssim $
0.5 fm will contribute.

\section{IOPE and sum rules}

Our study will be based on the correlation function 
\begin{equation}
\Pi \left( -q^{2}\right) =i\int d^{4}xe^{iqx}\left\langle 0|T\,O_{S}\left(
x\right) O_{S}\left( 0\right) |0\right\rangle   \label{corr}
\end{equation}
with the interpolating field 
\begin{equation}
O_{S}=\alpha _{s}G_{\mu \nu }^{a}G^{a,\mu \nu }  \label{intpol}
\end{equation}
carrying the quantum numbers of the scalar glueball. The standard OPE of
this correlator, including perturbative Wilson coefficients up to $O\left(
\alpha _{s}\right) $ and operator contributions up to dimension 8, is known
to be \cite{bag90} (up to polynomials in $Q^{2}$) 
\begin{align}
\Pi ^{\left( OPE\right) }(Q^{2})& =Q^{4}\ln \left( \frac{Q^{2}}{\mu ^{2}}%
\right) \left[ -2\left( \frac{\alpha _{s}}{\pi }\right) ^{2}\left[ 1+\frac{59%
}{4}\frac{\alpha _{s}}{\pi }\right] +b_{0}\left( \frac{\alpha _{s}}{\pi }%
\right) ^{3}\ln \left( \frac{Q^{2}}{\mu ^{2}}\right) \right]   \nonumber \\
& +4\alpha _{s}\left[ 1+\frac{49}{12}\frac{\alpha _{s}}{\pi }\right]
\left\langle \alpha _{s}G^{2}\right\rangle -\frac{\alpha _{s}^{2}b_{0}}{\pi }%
\left\langle \alpha _{s}G^{2}\right\rangle \ln \left( \frac{Q^{2}}{\mu ^{2}}%
\right)   \nonumber \\
& +\frac{1}{Q^{2}}\left[ 8\alpha _{s}^{2}\left\langle gG^{3}\right\rangle
L^{7/11}-58\alpha _{s}^{3}\left\langle gG^{3}\right\rangle L^{7/11}\ln
\left( \frac{Q^{2}}{\mu ^{2}}\right) \right] +8\pi \alpha _{s}\frac{1}{Q^{4}}%
\left\langle \alpha _{s}^{2}G^{4}\right\rangle 
\end{align}
($Q^{2}=-q^{2}$) where $b_{0}=11N_{c}/3-2N_{f}/3$ is the leading-order
contribution to the QCD $\beta $ function, $L\left( Q^{2}\right) =\ln \left(
Q/\Lambda \right) /$ $\ln \left( \mu _{0}/\Lambda \right) $, and $\alpha _{s}
$ is the QCD running coupling at one loop. We will use the parameter values $%
\Lambda =0.12$ GeV, $\mu =0.5$ GeV and the condensate values $\left\langle
\alpha _{s}G^{2}\right\rangle =0.04$ GeV$^{4}$, $\left\langle
gG^{3}\right\rangle =-1.5\left\langle \alpha _{s}G^{2}\right\rangle ^{3/2}$
of Ref. \cite{bag90}. For the four-gluon condensate, finally, we adopt the
standard approximation 
\begin{equation}
\left\langle \alpha _{s}^{2}G^{4}\right\rangle \equiv 14\left\langle (\alpha
_{s}f_{abc}G_{\mu \rho }^{b}G_{\nu }^{\rho c})^{2}\right\rangle
-\left\langle (\alpha _{s}f_{abc}G_{\mu \nu }^{b}G_{\rho \lambda
}^{c})^{2}\right\rangle \simeq \frac{9}{16}\left\langle \alpha
_{s}G^{2}\right\rangle ^{2}.
\end{equation}

In addition to the perturbative contributions given above, the Wilson
coefficients receive nonperturbative contributions from direct instantons
which have so far been neglected in the gluonium sum rules (a partial
estimate was given in Ref. \cite{shu82}). Analogous contributions were found
to be important in several nucleon \cite{for93} and pion \cite{shu83}\ sum
rules. As noted there, effects of multi-instanton correlations can be
neglected in the short-distance expansion since the relevant distances $%
\left| x\right| \sim \left| Q^{-1}\right| \leq 0.2$ fm are much smaller than
the average separation $\bar{R}\sim 1$ fm between instantons in the vacuum.
The diluteness of the instanton vacuum distribution, which is a consequence
of $\bar{\rho}/\bar{R}\ll 1$, further reduces the impact of multi-instanton
correlations and keeps the separate instantons approximately undeformed.

To leading order in the semiclassical approximation, the instanton
contribution to Eq. (\ref{corr}) can thus be calculated by standard
techniques \cite{nov280} from the $O\left( \hbar ^{0}\right) $ component of
the gluon propagator in the instanton background \cite{bro78} and reads 
\begin{equation}
\Pi ^{\left( I+\bar{I}\right) }\left( Q^{2}\right) =2^{5}\pi ^{2}\bar{n}\bar{%
\rho}^{4}Q^{4}K_{2}^{2}\left( Q\bar{\rho}\right) .  \label{icor1}
\end{equation}
($K_{2}$ is a McDonald function.) Since the average instanton size $\bar{\rho%
}\simeq \left( 1/3\right) $ fm is small compared to $\Lambda _{QCD}^{-1}$, $%
O\left( \hbar \right) $ corrections to Eq. (\ref{icor1}) are suppressed by
the large instanton action $S_{I}\left( \bar{\rho}\right) \sim 10\hbar $.
Instanton contributions to the Wilson coefficients of power corrections
carry additional inverse powers of the relatively large glueball mass scale
and are therefore also expected to be small (see Ref. \cite{for00} for a
more detailed discussion).

Various sum rules can be constructed from the Borel transform of weighted
moments of the glueball correlator, 
\begin{equation}
{\cal R}_{k}\left( \tau \right) =\hat{B}\left[ \left( -Q^{2}\right) ^{k}\Pi
\left( Q^{2}\right) \right] .  \label{momdef}
\end{equation}
Typically, one considers $k\in \left\{ -1,0,1,2\right\} $. The corresponding
expressions for ${\cal R}_{k}^{\left( OPE\right) }$ (together with the
explicit form of the Borel operator $\hat{B}$) are given in Ref. \cite{bag90}%
. The instanton contributions ${\cal R}_{k}^{\left( I+\bar{I}\right) }$ are
obtained recursively, via 
\begin{equation}
{\cal R}_{k}^{\left( I+\bar{I}\right) }\left( \tau \right) =\left( \frac{%
-\partial }{\partial \tau }\right) ^{k+1}{\cal R}_{-1}^{\left( I+\bar{I}%
\right) }\left( \tau \right) \text{\ \ \ \ \ \ \ \ }\left( k\geq -1\right) 
\label{ri}
\end{equation}
from ${\cal R}_{-1}^{\left( I+\bar{I}\right) }$, which can be calculated in
closed form [$x\equiv \bar{\rho}^{2}/\left( 2\tau \right) $]: 
\begin{equation}
{\cal R}_{-1}^{\left( I+\bar{I}\right) }\left( \tau \right) =-2^{6}\pi ^{2}%
\bar{n}x^{2}e^{-x}\left[ \left( 1+x\right) K_{0}\left( x\right) +\left( 2+x+%
\frac{2}{x}\right) K_{1}\left( x\right) \right] +2^{7}\pi ^{2}\bar{n}.
\label{instpart}
\end{equation}

The sum ${\cal R}_{k}^{\left( OPE\right) }+{\cal R}_{k}^{\left( I+\bar{I}%
\right) }$ constitutes the IOPE. Note that we have removed the constant
subtraction term $-\Pi ^{\left( I+\bar{I}\right) }\left( 0\right) =-2^{7}\pi
^{2}\bar{n}$ in Eq. (\ref{instpart}) because it originates from soft
instanton contributions which do not belong to the OPE coefficients.
Double-counting of soft instanton physics is thereby excluded since the
instanton contributions (\ref{ri}) do not contain powers $\tau ^{n}$ with $%
n>-\left( k+3\right) $.

In order to write down the sum rules, we have to match the IOPE expressions
to their ``phenomenological'' counterparts, which are derived from the twice
subtracted dispersion relation 
\begin{equation}
\Pi ^{\left( phen\right) }\left( Q^{2}\right) =\Pi ^{\left( phen\right)
}\left( 0\right) -\Pi ^{\left( phen\right) ^{\prime }}\left( 0\right) Q^{2}+%
\frac{\left( Q^{2}\right) ^{2}}{\pi }\int_{0}^{\infty }ds\frac{%
%TCIMACRO{\func{Im}}%
%BeginExpansion
\mathop{\rm Im}%
%EndExpansion
\Pi ^{\left( phen\right) }\left( s\right) }{s^{2}\left( s+Q^{2}\right) }
\label{disprel}
\end{equation}
by parametrizing the spectral function in terms of a pole contribution and
an effective continuum. Following standard procedure, the latter is obtained
from the dispersive cut of the theoretical side and starts at an effective
threshold $s_{0}$. Thus we have 
\begin{equation}
%TCIMACRO{\func{Im}}%
%BeginExpansion
\mathop{\rm Im}%
%EndExpansion
\Pi ^{\left( phen\right) }\left( s\right) =\pi f_{G}^{2}m_{G}^{4}\delta
\left( s-m_{G}^{2}\right) +\left[ 
%TCIMACRO{\func{Im}}%
%BeginExpansion
\mathop{\rm Im}%
%EndExpansion
\Pi ^{\left( OPE\right) }\left( s\right) +%
%TCIMACRO{\func{Im}}%
%BeginExpansion
\mathop{\rm Im}%
%EndExpansion
\Pi ^{\left( I+\bar{I}\right) }\left( s\right) \right] \theta \left(
s-s_{0}\right) .  \label{specdens}
\end{equation}

As a consequence of the exceptional size of the instanton contributions to
the scalar glueball correlator, their contributions to the continuum are an
indispensable part of Eq. (\ref{specdens}) and will turn out to play an
essential role in the subsequent analysis. Explicitly, we find 
\begin{equation}
%TCIMACRO{\func{Im}}%
%BeginExpansion
\mathop{\rm Im}%
%EndExpansion
\Pi ^{\left( I+\bar{I}\right) }\left( s\right) =-2^{4}\pi ^{4}\bar{n}\bar{%
\rho}^{4}s^{2}J_{2}\left( \sqrt{s}\bar{\rho}\right) Y_{2}\left( \sqrt{s}\bar{%
\rho}\right) 
\end{equation}
where the $J_{2}$ ($Y_{2}$) are Bessel (Neumann) functions. (A more detailed
multipole analysis, allowing for neighboring and mixed quarkonium
resonances, will be relegated to Ref. \cite{for00}.)

By equating the phenomenological Borel moments, obtained from Eqs. (\ref
{momdef}) and (\ref{disprel}), to the corresponding IOPE expressions one
finally obtains the sum rules 
\begin{equation}
\frac{{\cal R}_{k}\left( \tau ,s_{0}\right) }{m_{G}^{2+2k}}%
=f_{G}^{2}m_{G}^{2}e^{-\tau m_{G}^{2}}  \label{sr}
\end{equation}
with 
\begin{equation}
{\cal R}_{k}\left( \tau ,s_{0}\right) =\sum_{X=OPE,I+\bar{I}}\left[ {\cal R}%
_{k}^{\left( X\right) }\left( \tau \right) -{\cal R}_{k}^{\left(
X-cont\right) }\left( \tau ,s_{0}\right) \right] +\delta _{k,-1}\Pi ^{\left(
phen\right) }\left( 0\right)
\end{equation}
and 
\begin{equation}
{\cal R}_{k}^{\left( X-cont\right) }\left( \tau ,s_{0}\right) =\frac{1}{\pi }%
\int_{s_{0}}^{\infty }dss^{k}%
%TCIMACRO{\func{Im}}%
%BeginExpansion
\mathop{\rm Im}%
%EndExpansion
\Pi ^{\left( X\right) }\left( s\right) e^{-s\tau }.
\end{equation}

Note that the higher moments weight the higher-mass region of the spectral
function more strongly and thus receive enhanced contributions from the
relatively heavy (see below) glueball pole. The subtraction constant $\Pi
^{\left( phen\right) }\left( 0\right) $ in\ the ${\cal R}_{-1}$ sum rule
(regularized by removing the high-momentum contributions) can be related to
the gluon condensate by the low-energy theorem (LET) \cite{nov280} 
\begin{equation}
\Pi \left( 0\right) =\frac{32\pi }{b_{0}}\left\langle \alpha
G^{2}\right\rangle .  \label{let}
\end{equation}
This relation provides an important consistency check for the sum-rule
results, as we will discuss below.

\section{Sum rule analysis}

The quantitative analysis of the sum rules amounts to determining those
values of the hadronic parameters in Eq. (\ref{specdens}) for which both
sides of Eq. (\ref{sr}) optimally match in the fiducial Borel domain.
Towards large $\tau $ this domain is bounded by keeping the contribution of
the highest-dimensional operator ($\alpha _{s}^{2}G^{4}$) to ${\cal R}_{k}$
below 10\% and requiring multi-instanton contributions to be negligible. The
latter requirement will be (conservatively) implemented by demanding $\tau
\leq 1$ GeV$^{-2}$.\ Towards small $\tau $ we prescribe that the continuum
contributions do not exceed 50\% of the ${\cal R}_{k}$.

The standard optimization procedure followed in previous analyses determined
only the glueball mass $m_{G}$ and coupling $f_{G}$ from matching the sum
rules, while the threshold $s_{0}$ had to be found by other means (e.g. by
finite-energy sum rules \cite{bag90} or stability criteria \cite{nar98}).
The IOPE sum rules turn out to be stable enough, however, to determine $%
s_{0} $ together with the resonance parameters $m_{G}$ and $f_{G}$ from the
same sum rule. This is the procedure which we will adopt below.

We start by analyzing the ${\cal R}_{0}$ sum rule [i.e., Eq. (\ref{sr}) with 
$k=0$]. Figure 1 shows both sides\ of the optimized sum rule and separately
the three components (OPE with subtracted OPE continuum, instanton
contribution, and instanton continuum) which make up its left-hand side. The
matching between both sides of the sum rule is almost perfect over the whole
fiducial region. Comparing standard OPE and instanton (including continuum)
contributions shows that the latter are about 5 times larger. Thus, the
instanton contributions strongly dominate over the whole fiducial region and
increase the predictions for $f_{G}^{2}$ by about a factor of 5, resulting
in $f_{G}=1.14$ GeV. A similarly strong enhancement of $f_{G}$ was found in
instanton vacuum model calculations \cite{sch95}. The prediction for the
glueball mass, $m_{G}=1.40$ GeV, on the other hand, differs surprisingly
little from what is obtained without the instanton part. The continuum
threshold becomes $s_{0}=5.1$ GeV$^{2}$.

Figure 1 furthermore reveals that the instanton contributions to the
unitarity cut are indispensable for generating an exponential $\tau $
behavior below $\tau \simeq 0.8$ GeV$^{-2}$, and thus for an acceptable fit
to the pole contribution. (For larger values of $\tau $ the continuum
contributions are practically negligible and a fit to the instanton
contribution alone would become possible, although mostly outside of the
fiducial domain and with about 20\% smaller values for $m_{G}$ and $f_{G}^{2}
$.) It should also be noted that the hard nonperturbative instanton physics
begins to enter ${\cal R}_{0}$ at much smaller $\tau $ than the soft
condensate contributions, thereby confirming the existence of an
exceptionally large mass scale in the scalar glueball channel \cite{nov81}.

The analysis of the remaining sum rules (which all show a high degree of
stability) confirms the above observations about the role of the instanton
contributions. As an additional example, we plot in Fig. 2 the separate
contributions to the ${\cal R}_{2}$ sum rule and the fit of both sides,
which is again excellent. Since the instanton contribution is somewhat less
pronounced than in ${\cal R}_{0}$, we find a smaller value for the coupling: 
$f_{G}=1.01$ GeV. The result for the glueball mass increases to $m_{G}=1.53$
GeV, and the threshold $s_{0}=4.89$ GeV$^{2}$ is slightly reduced. The
results of the ${\cal R}_{-1}$ and ${\cal R}_{1}$ sum rules confirm the
tendency of lower moments to predict somewhat smaller masses and somewhat
larger couplings [while maintaining consistency with the low-energy theorem (%
\ref{let})]. The predictions of the higher moments should be more reliable,
however, because they receive stronger pole contributions.

The ${\cal R}_{-1}$ sum rule has played both conceptually and practically a
special role since it contains the subtraction constant $\Pi \left( 0\right) 
$ which dominates the power corrections of the conventional OPE (in the
fiducial region). Attempts to fit the resulting, almost flat $\tau $
behavior to the exponential pole contribution inevitably generate very small
pole masses, at least half an order of magnitude smaller than those
predicted by the other sum rules. This well-known inconsistency (and the
need to abandon the ${\cal R}_{-1}$ sum rule in practice) is largely
resolved by the massive instanton contributions. Their strong decay yields
excellent fits and pole masses of the same order as those obtained from the
higher moments. Still, the ${\cal R}_{-1}$ sum rule remains probably least
suited for quantitative predictions since it is most sensitive to the
inaccurately known value of the gluon condensate and least sensitive to the
pole contribution. The mutual agreement of all four IOPE sum rules (in the
typical range of uncertainty for sum rule results) and their consistency
with the low-energy theorem\footnote{%
We have checked that the zero-momentum correlator $\Pi ^{\left( reg\right)
}\left( 0\right) $, obtained from the UV-regularized, unsubtracted
dispersion relation with the spectral density (\ref{specdens}) (where $m_{G}$%
, $f_{G}$, and $s_{0}$ have the predicted values) \ satisfies the low-energy
theorem (\ref{let}) for all four sum rules in the range of uncertainty
introduced by the inaccurately known value of the gluon condensate \cite
{for00}.}, however, is reassuring and of considerable conceptual importance.

For a quantitative consistency check between the predictions of different
sum rules, we have evaluated the $\tau $-dependent mass function 
\begin{equation}
m_{G}^{\left( 1,2\right) }\left( \tau \right) \equiv \sqrt{\frac{{\cal R}%
_{2}\left( \tau ,s_{0}\right) }{{\cal R}_{1}\left( \tau ,s_{0}\right) }}.
\end{equation}
Figure 3 shows that it deviates less than 2\% from the constant $m_{G}$ over
the whole fiducial region, indicating a high degree of compatibility between
the sum rules.

We have also found that the instanton contributions to the ${\cal R}_{k}$ by
themselves can generate stable sum rules. Their approximately exponential $%
\tau $ behavior matches very well to the pole term, although with about 20\%
smaller glueball masses than those obtained from the full sum rules. This
indicates that instantons alone can (over-) bind the scalar glueball, in
agreement with the findings of instanton vacuum models \cite{sch95} (which
also show a tendency towards smaller glueball masses).

\section{Discussion and conclusions}

We evaluated and analyzed the instanton contributions both to the OPE of the
scalar glueball correlator (or, more precisely, to the Wilson coefficient of
the unit operator) and to the continuum part of its phenomenological
spectral-function model, and we solved the corresponding QCD sum rules. The
previously neglected instanton contributions turn out to be dominant and
render the IOPE sum rules the first overall consistent set in the scalar
glueball channel.

In particular, the IOPE resolves two long-standing flaws of the earlier sum
rules: the mutual inconsistency between different Borel moments and the
inconsistency with the low-energy theorem for the zero-momentum correlator.
Even the previously deficient and usually discarded lowest-moment (${\cal R}%
_{-1}$) sum rule becomes consistent both with those from the higher moments
and with the low-energy theorem. Any evidence for a low-lying ($m\ll 1$ GeV)
gluonium state (or a state strongly coupled to gluonic interpolators),
sometimes argued for on the basis of this sum rule \cite{nov280,nov81}, is
thereby rendered obsolete.

The most dramatic phenomenological impact of the direct instanton
contributions is associated with $f_{G}^{2}$, the residuum of the glueball
pole. Due to the exceptional size of the instanton contributions, its value
increases by about a factor of 5. Taking the quantitative predictions of the 
${\cal R}_{2}$ sum rule to be the most reliable ones, we obtain $%
m_{G}=1.53\pm 0.2$ GeV (in accord with recent lattice results \cite{lat})
and $f_{G}=1.01\pm 0.25$ GeV, where the errors are estimated from the
uncertainties of the input parameters and the spread between the individual
sum rules. Potential ramifications for experimental glueball searches will
be considered in a forthcoming publication.

All four IOPE sum rules show an unprecedented degree of stability and allow
for a simultaneous 3-parameter fit to the glueball mass, its coupling, and
the continuum threshold. The stability region extends far beyond the
fiducial $\tau $ interval and renders, as a side effect, the IOPE sum rule
results almost insensitive to the precise boundaries of the fiducial domain.
Most importantly, however, the high stability indicates that the IOPE
provides a rather complete description of the short-distance glueball
correlator.

A crucial contribution to the IOPE sum rules arises from the discontinuity
of the instanton terms in the extended continuum part of the spectral
functions. (The rough estimate of instanton contributions to the ${\cal R}%
_{0}$ sum rule in Ref. \cite{shu82} missed this contribution.) In addition
to substantially improving the overall consistency and stability of the sum
rules, the richer structure on the phenomenological sides also sheds new
light on the spectral content of the scalar glueball correlator. For once,
the quantitative sum-rule analysis reveals that the instanton contributions,
together with the weaker perturbative terms, counterbalance the pole
contribution. This leads to an improved description of the correlator
towards low momenta and thereby reconciles the sum rules with the low-energy
theorem (\ref{let}), a stringent consistency check\footnote{%
On the lattice, the feasibility of this check is compromised by finite-size
effects.} in the $Q\rightarrow 0$ limit.

A remarkable interplay between perturbative and nonperturbative physics can
also be seen in the opposite limit, i.e. at short distances. As a
consequence of the improved continuum description, the instanton
contributions to the correlator remain effective at small $\tau $ and stay
finite even for $\tau \rightarrow 0$ (in contrast to their contributions to
the IOPE in the same limit, which suffer the expected suppression associated
with funnelling a sizeable momentum through the coherent instanton field).
This indicates that small-instanton physics accounts not only for much of
the ground-state contribution but also for part of the higher-lying spectral
strength (in the sense of a generalized quark-hadron duality) in the scalar
glueball correlator. Thus, the spectral distribution favored by the IOPE\
sum rules seems to imply a rather prominent role for instanton-induced
effects in excited glueball states (or in multiparticle states with strong
coupling to the energy-momentum tensor). The phenomenological impact of
these results, which do not depend on details of the IOPE and should
therefore be rather robust, deserves further study.

In contrast to previously studied IOPE sum rules for quark-based correlators 
\cite{for93,shu83}, those for the scalar glueball are the first where (i)
the instanton contributions do not enter via topological quark zero-modes
and (ii) the sum rules reach a satisfactory (though not excellent) level of
consistency even without any perturbative and soft contributions, i.e., with
the instanton terms alone. The latter result explains why the instanton
liquid model yields scalar glueball properties similar to those obtained
above \cite{sch95}, and can be traced to both the exceptional strength and
the particular shape (mainly the curvature) of the instanton contributions.
In combination, those properties produce an approximately exponential $\tau $
dependence which extends far beyond the fiducial region and fairly well
matches the ground-state signal without any perturbative and condensate
contributions.

The predominance and approximate self-sufficiency of the instanton
contributions has a suggestive physical interpretation: it indicates that
instantons may generate the bulk of the attractive forces which bind the
scalar glueball. Moreover, and in contrast to the instanton liquid model of
Ref. \cite{sch95}, the IOPE allows to consistently compare the instanton
contributions with those of the remaining soft and perturbative fields, and
to thereby judge their relative importance. At present, the IOPE seems to be
the only controlled and analytical framework in which this can be achieved.
The combined effect of the soft and perturbative contributions turns out to
be repulsive and increases, consistently in all sum rules, the glueball mass
by about 20\%.

Finally, closer inspection of the IOPE\ sum rules reveals another, quite
striking instanton effect: the scales of the predicted $0^{++}$ glueball
properties turn out to be approximately set by the bulk features of the
instanton size distribution. Indeed, neglecting the standard OPE
contributions one finds that the glueball parameters scale as\footnote{%
It is worth noting that the arguments leading to these scaling relations
take advantage of the analytical character of the IOPE sum rules. The
scaling would be more difficult to uncover in numerical approaches (as in
lattice simulations, where in addition the instanton distribution parameters
cannot easily be varied).} 
\begin{eqnarray}
m_{G} &\sim &\bar{\rho}^{-1}, \\
f_{G}^{2} &\sim &\bar{n}\bar{\rho}^{2}.
\end{eqnarray}
In the case of proportionality between the glueball size $r_{G}$ and its
Compton wavelength, one would obtain another scaling relation 
\begin{equation}
r_{G}\sim \bar{\rho},
\end{equation}
which could explain the small values $r_{G}\sim $ 0.2 fm found on the
lattice \cite{def92}.

Conceptually, the main virtue of the above scaling relations lies in
establishing an explicit link between fundamental vacuum and hadron
properties. Although strong interdependences between QCD vacuum and hadron
structure are expected on general grounds, such scaling relations seem to
have not been encountered previously. They could be of practical use e.g.
for the test of instanton vacuum models, to provide constraints for glueball
model building, or generalized to finite temperature and baryon density,
where the scales of the instanton distribution change.

This work was supported by Deutsche Forschungsgemeinschaft under
Habilitation Grant Fo 156/2-1.

\section{Figure captions}

\begin{enumerate}
\item  Fig. 1:\ The right-hand side $f_{G}^{2}m_{G}^{2}\exp \left(
-m_{G}^{2}\tau \right) $ of the ${\cal R}_{0}$ sum rule (dotted), compared
with the optimized left-hand side ${\cal R}_{0}\left( \tau ,s_{0}\right)
/m_{G}^{2}$ (solid line) and its three components (all in units of GeV$^{4}$%
): the conventional OPE ${\cal R}_{0}^{\left( OPE\right) }\left( \tau
,s_{0}\right) /m_{G}^{2}$ (dash-double-dotted), the instanton contribution $%
{\cal R}_{0}^{\left( I+\bar{I}\right) }\left( \tau \right) /m_{G}^{2}$
(dashed), and the instanton continuum part $-{\cal R}_{0}^{\left(
I-cont\right) }\left( \tau ,s_{0}\right) /m_{G}^{2}$ (dash-dotted).

\item  Fig. 2: Same as Fig. 1 for the ${\cal R}_{2}$ sum rule.

\item  Fig. 3: The square root of the ratio ${\cal R}_{2}\left( \tau \right)
/{\cal R}_{1}\left( \tau \right) $. The weak $\tau $ dependence confirms the
high consistency between the ${\cal R}_{1}\left( \tau \right) $ and ${\cal R}%
_{2}\left( \tau \right) $ sum rules.
\end{enumerate}

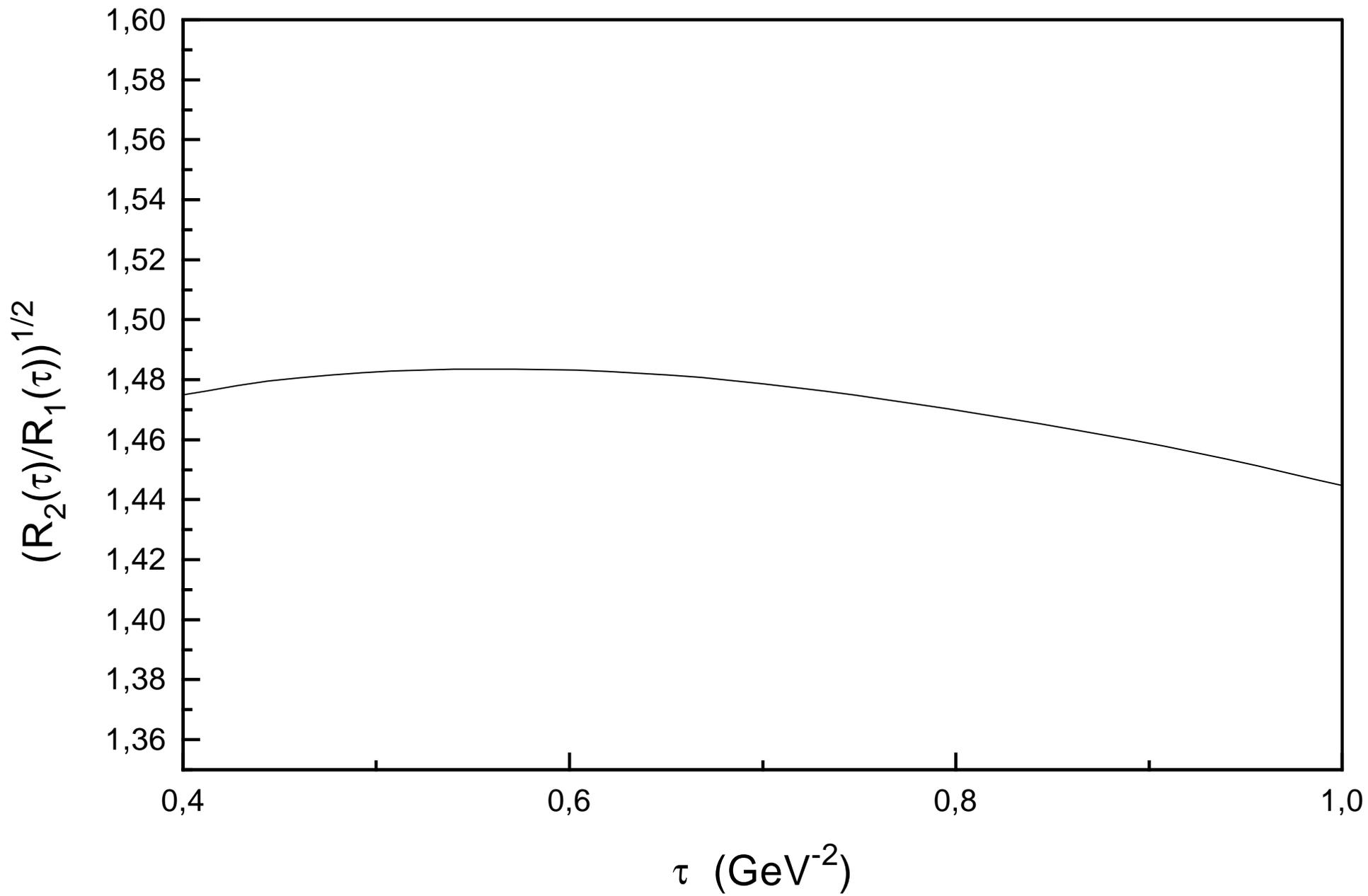
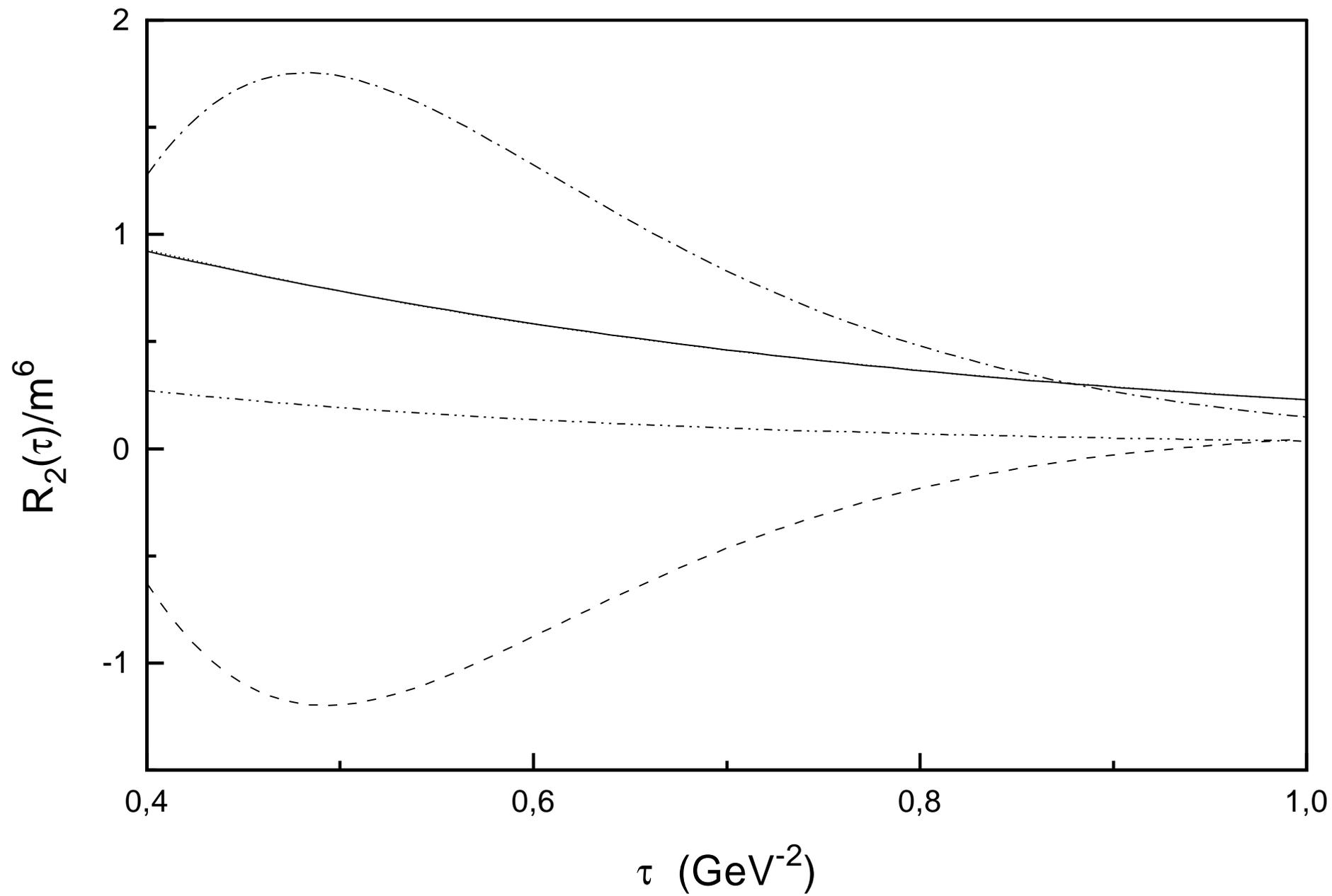
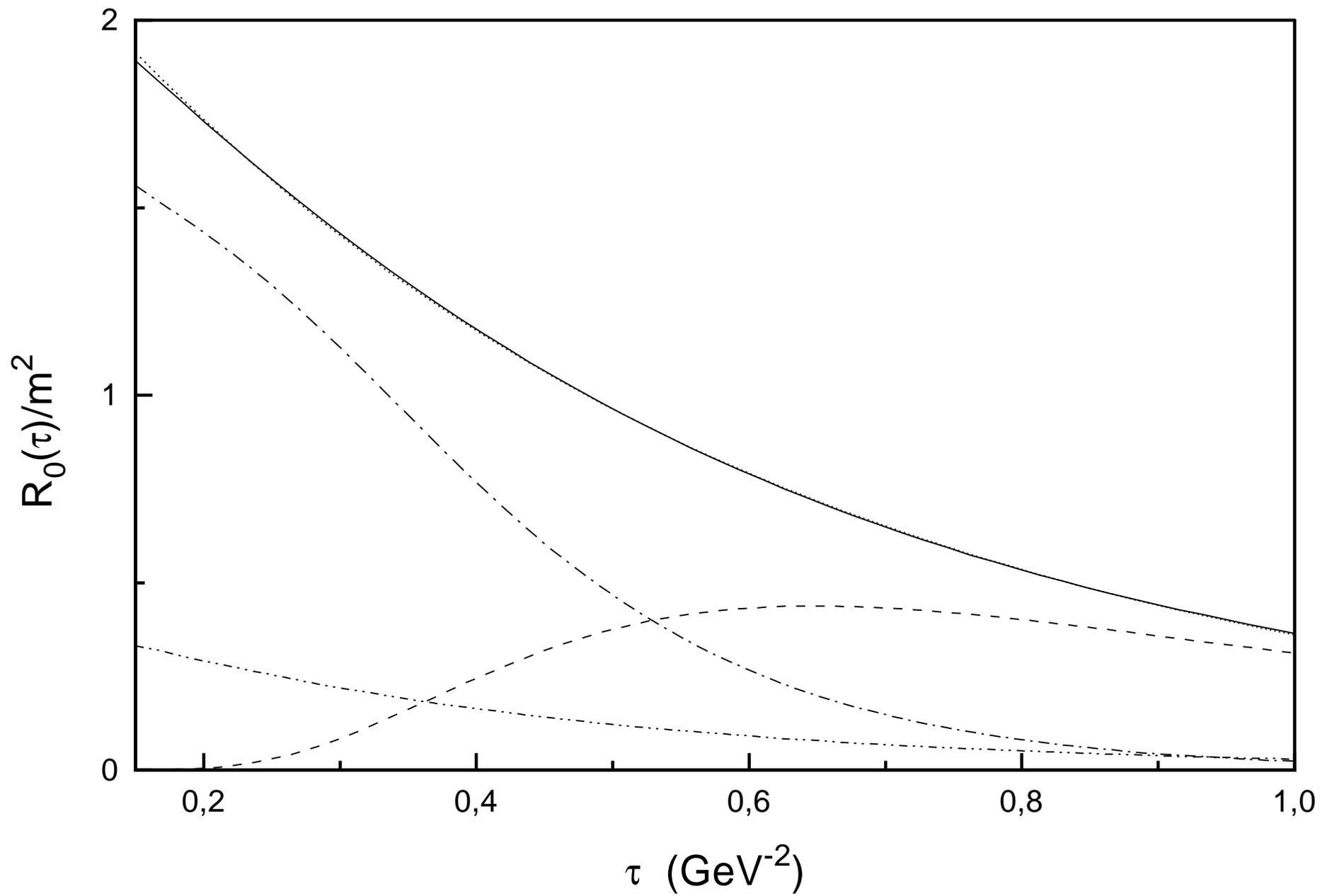
\end{document}